\documentclass[aps,showpacs,twocolumn,amsmath,amssymb]{revtex4}

\usepackage{graphicx}

\def\Eq#1{Eq.~(\ref{#1})}
\def\Eqs#1{Eqs.~(\ref{#1})}
\def\Fig#1{Fig.~\ref{#1}}
\def\Figs#1{Figs.~\ref{#1}}

\def\no{\nonumber\\}
\def\r{\rangle\!\rangle}
\def\l{\langle\!\langle}
\def\>{\rangle}
\def\<{\langle}
\def\d{\partial}
\def\phiv{\boldsymbol{\phi}}
\def\vphiv{\boldsymbol{\varphi}}
\def\psiv{\boldsymbol{\psi}}
\def\chiv{\boldsymbol{\chi}}
\def\fb{\boldsymbol{f}}
\def\gb{\boldsymbol{g}}
\def\Khat{\hat{\mathcal{K}}}
\def\K{\boldsymbol{K}}
\def\Tb{\bar{T}}
\def\tb{\bar{t}}
\def\btb{\bar{\beta}}
\def\pb{\bar{p}}
\def\p0{{\bar{p}_0}}
\def\phit{\tilde{\phi}}
\def\phip{\phi^+}
\def\phim{\phi^-}

\def\dg{\dagger}

\def\Iii{\int_{-\infty}^{\infty}}

\def\half{\tfrac{1}{2}}
\def\lb{\bar{\lambda}}
\def\lam{\lambda}
\def\ome{\omega}
\def\del{\delta}

\begin{document}

\title{Band Structure and Accumulation Point in the Spectrum of Quantum Collision Operator in a One-Dimensional Molecular Chain}

\author{B. A. Tay,$^1$\footnote{Email:batay@um.edu.my} Kazuki Kanki,$^2$ Satoshi Tanaka$^2$ and Tomio Petrosky$^3$}
\affiliation{
$^1$Department of Physics, Faculty of Science, University of Malaya, 50603 Kuala Lumpur, Malaysia\\
$^2$Department of Physical Science, Osaka Prefecture University, Sakai, Osaka 599-8531, Japan\\
$^3$Center for Complex Quantum Systems,
University of Texas at Austin, Texas 78712, USA}

\date{\today}

\begin{abstract}
We consider the eigenvalue problem of a kinetic  collision operator for a quantum Brownian particle interacting with a one-dimensional chain. The quantum nature of the system gives rise to a difference  operator. For the one-dimensional case, the momentum space separates into infinite sets of disjoint subspaces dynamically independent of one another. The eigenvalue problem of the collision operator is solved with the continued fraction method. The spectrum is non-negative, possesses an accumulation point and exhibits a band structure. We also construct the eigenvectors of the collision operator and establish their completeness and orthogonality relations in each momentum subspaces.

\end{abstract}


\maketitle

\section{Introduction}

Since the seminal work of Boltzmann's on dilute gas \cite{Boltzmann}, the kinetic equation approach had been extensively applied to various fields, for example, the transport phenomena in solid state physics \cite{Peierls55,Mahan55}, dense fluids and plasma \cite{Balescu,Resibois}, optical transition in quantum optics and quantum information \cite{Carmichael,Nielson}, the motion of quantum Brownian particle in a potential \cite{Risken}, and etc. The kinetic equation incorporates the effect of fluctuation on a test particle by considering the collision between the particle with its environment.

A density matrix  formulation for the relaxation process is desired since dissipation inevitably brings a pure state into a mixed state, which is incompatible with the unitary time evolution on the wave function level through the Schr\"odinger equation. This results in a non-hermitian collision operator that governs the relaxation of the density matrix of the particle towards equilibrium. It is therefore important to study the eigenvalue problem of the collision operator to understand the relaxation process.

As has been shown in Ref.\cite{Petrosky97}, the eigenvalue problem of the collision operator is closely related with the eigenvalue problem of the Liouville operator.  However, spectral properties of the Liouville operator, as well as spectral properties of the collision operator, are not yet well-analyzed as compared with the spectral properties of the Hamiltonian.  Indeed, despite the versatility of the kinetic equation approach and the existence of various methods in studying it \cite{Balescu,Resibois,Risken}, only a few models are known to be exactly solvable, such as, e.g.,  the Maxwell molecules\cite{Balescu} and the particle-field linear coupling model  in one dimension (1D)\cite{Agarwal}. The model that we are going to study is another example of an exactly soluble model in 1D on the level of kinetic equation. It gives rise to a few interesting results to be discussed below.

We study a quantum Brownian particle interacting with a 1D chain. The model is derivable, for instance, from an electron interacting with a background 1D ion lattice through a deformation potential \cite{Bardeen50,Frochlich52}, or a Davydov soliton produced by peptide oscillation propagating under the influence of a 1D molecular chain \cite{Davydov}. Previous studies on this model were mainly carried out on the wave function level \cite{Davydov}. Here, we focus on the relaxation process on the level of momentum distribution function of the particle under the thermal fluctuation of the background 1D chain.

The collision operator of this system exhibits a few interesting features \cite{Tanaka}. In our quantum system, it takes a  form of a difference operator, in sharp contrast to the usual differential operator in classical systems, as well as linear coupling quantum systems. Moreover, the dissipative effect introduced by the collision operator is a purely quantum effect similarly to the case of the kinetic equation in 1D classical gas.  Our collision operator possesses also a special character owing to its 1D nature, i.e., the momentum states are coupled discretely, resulting in infinite sets of dynamically disjoint momentum subspaces.

The difference equation in the momentum space renders the usual method in solving differential equations (such as eigenfunction expansion and the Green's function method) inapplicable to this case. We instead find that the eigenvalue problem is best solved with the continued fraction method \cite{Morse,Risken}, which can be efficiently implemented numerically.

The main results of the paper are as follows. We find that the spectrum of the difference operator is discrete in each disjoint momentum subspace and non-negative. It possesses a maximum eigenvalue in each momentum subspaces and an accumulation point. Moreover, the eigenvalues are non-degenerate except in the steady mode, thus giving rise to a band structure.

In Ref.~\cite{Kanki10}, we presented some of the main features of the solution by continued fraction method and the spectrum, but without much discussion on the mathematical details. Here, we clarify the mathematical structure of the eigenvalue problem of the collision operator. While Ref.~\cite{Kanki10} used a symmetrized version of the collision operator, here we work with the original non-hermitian collision operator. We clarify the biorthogonal space structure \cite{Morse} of the operator that belongs to the generalized space \cite{gensp}. We also introduce a scalar product to the Liouville space, clarify the detailed structure of the spectrum and construct the eigenvectors.

The discussion is organized as follows. In Section \ref{sec-mod} we introduce the model and the difference collision operator. The disconnectedness nature of the momentum space is also explained. In Section \ref{sec-eigcoll} we consider the eigenvalue problem and clarify the biorthogonal structure of the eigenvector. We then use the continued fraction method to solve the eigenvalue problem in Section \ref{sect-eigenvalue} and discuss the details of the spectrum in Section \ref{sec-spectrum}. In Section \ref{sec-approx} we construct the eigenfunctions of the collision operator and make a comparison with the approximate solution obtained by truncating the difference collision operator to be a second order differential operator. In Section \ref{Concl} we present conclusions.

\section{Difference collision operator}
\label{sec-mod}

\subsection{Model}

We consider the relaxation dynamics of a quantum Brownian particle interacting with the acoustic phonon field of a 1D lattice \cite{Frochlich52,Bardeen50},
\begin{align}
H&=\sum_p \varepsilon_p |p\rangle\langle p|
 +\sum_q\hbar\omega_q a_q^\dagger a_q \notag\\
 &+\sqrt{\frac{2\pi}{L}} \sum_{p,q} V_q |p+\hbar q\rangle\langle p|
  (a_q + a_{-q}^\dagger)\,,
\end{align}
where we restrict our consideration to the one particle sector of the Brownian particle. The momentum state vector of the particle is labeled by $|p\rangle$. The creation and annihilation operators of the phonon field with wave vector $q$ are $a_q^\dg$ and $a_q$, respectively. The length $L$ is the dimension of the lattice. The particle has mass $m$ and energy $\varepsilon_p=p^2/2m$, whereas the acoustic phonons assume the dispersion relation $\omega_q=c|q|$. The particle is coupled to the phonon field through a deformation potential with effective coupling \cite{Mahan55,Whitfield}
\begin{align}   \label{deformpot}
   V_q=\sqrt{\frac{\hbar\Delta_0^2 |q|^2}{4\pi\rho_M\omega_q}}\,,
\end{align}
in which $\Delta_0$ is the deformation potential and
$\rho_M $ is the mass density of the chain.
We consider a weakly coupling case between the particle and phonons.
We impose a periodic boundary condition
leading to discrete
momentum $p$ and wave numbers $q$ with  $ p/\hbar, q = 2\pi j/L$, where $j=0, \pm 1, \pm 2, \cdots$, respectively.

The time evolution of the total system follows the Liouville-von Neumann equation,
\begin{equation}
i\frac{\partial}{\partial t}\rho(t)=\mathcal{L}\rho(t),
\end{equation}
with the  Liouvillian $\mathcal{L}$ defined by
\begin{equation}
\mathcal{L}\rho \equiv \frac{1}{\hbar} [H,\rho],
\end{equation}
where $\rho(t)$ is the density matrix of the total system,
and the Liouvillian $\mathcal{L}$ is defined by
$\mathcal{L}\rho=[H,\rho]/\hbar$.
We are interested in the time evolution
of the reduced density
matrix of the particle, and we average over the phonon field,
\begin{equation}
f(t)  \equiv  \mathrm{Tr}_\mathrm{ph}\rho(t).
\end{equation}
We assume that the phonon system is in thermal equilibrium with a temperature $T$ represented  by
\begin{equation}
\rho_{\rm ph}^{\rm eq} \propto \exp[ - \sum_q \beta  \hbar \ome_q a_q^\dagger a_q] ,
\end{equation}
where $\beta \equiv 1/k_B T$ with the Boltzmann constant $k_B$.

We are interested in the limit $ L \to \infty$, where the wave number becomes continuous and the summation over the momentum is replaced by an integration,
\begin{align}   \label{discint}
    \frac{2\pi}{L}\sum_p \to \int dp\,, \quad \frac{L}{2\pi} \del_{p,p'}\to \del(p-p') \,.
\end{align}
We will continue to use the discrete notation in our presentation for compactness of expression.

For the weak coupling system, the effect of the interaction between the particle and phonons can be approximated to the second order with respect to the potential $V_q$, and the time evolution equation of the momentum distribution function defined as $f(p,t) \equiv \< p |f(t) |p \>$ obeys a Markovian kinetic equation in a form $i\d f(p, t)/dt =\hat{\Psi}f(p, t)$, where $\hat{\Psi}$ is the generator of the time evolution (see e.g. Ref.\cite{Prigogine62}).

In view of the fact that the eigenvalues of $\hat{\Psi}$ are pure imaginary (which will be shown in due course),  we will  consider instead the operator $\Khat \equiv - i\hat{\Psi}$, where $\Khat $ is the {\it collision operator} that we will focus on from now on.

The reduced dynamics follows a Markovian kinetic equation \cite{Tanaka}
\begin{equation}\label{kinetic_eqq}
    \frac{\partial}{\partial t} f(p,t)= \Khat  f(p,t)\,,
\end{equation}
where the action of the collision operator on the distribution function is given by
\begin{align}\label{collop}
    &\Khat f(p,t)
    =
    - \frac{2\pi}{\hbar^2} \int dq|V_q|^2   \bigg\{ \del\left(\frac{\varepsilon_{p-\hbar q}-\varepsilon_p}{\hbar}+\ome_q\right) \no
         &\qquad \times  \bigg( [n(q)+1]f(p,t)-n(q)f(p-\hbar q,t)\bigg)
           \no
         &\qquad\qquad\qquad\qquad\qquad\quad+\del\left(
             \frac{\varepsilon_p-\varepsilon_{p+\hbar q}}{\hbar}+\ome_q \right) \no
         &\qquad \times   \bigg( n(q) f(p,t)-[n(q)+1] f(p+\hbar q,t)\bigg) \bigg\} \,.
\end{align}
The number density of the phonon obeys the Bose-Einstein distribution $n(\hbar q)=1/[\exp(\beta\hbar\omega_q)-1]$.
We note that as $\hbar$ approaches zero, the collision operator vanishes \cite{Tanaka}. Therefore, the dissipation caused by the collision operator is a purely quantum effect.

Notice that the collision operator takes the form of a difference operator. It can be expanded in an infinite series of higher derivative with respect to momentum, $f(p\pm\hbar q)=\exp(\pm\hbar q \d/\d p) f(p)$. This is in contrast to the usual stochastic differential kinetic equation, such as the Fokker-Planck equation \cite{Risken}, which contains differential operators up to the second order in momentum. We cannot in general approximate the difference kinetic equation by truncating the series of differential operators to more than the second order, otherwise the positivity of the reduced distribution function cannot be maintained \cite{Pawula,Risken}.

\subsection{Disjoint momentum subspaces}

\begin{figure}[htb]
  \centering
\includegraphics[width=2.in]{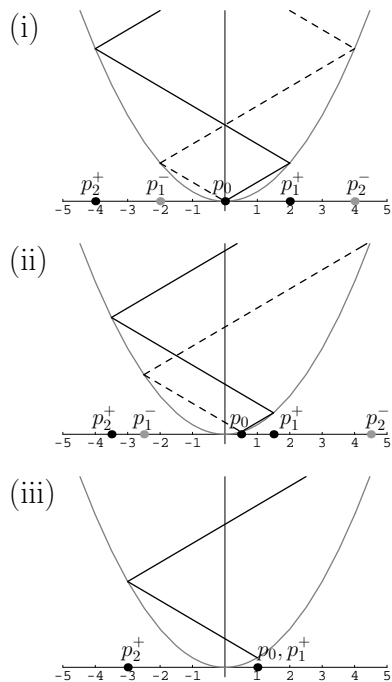}
\caption{Disjoint momentum subspaces, for (i) $\p0=0$, (ii) $-1<\p0<1$ and (iii) $\p0=1$. Solid lines connect the positive branch, and dotted lines connect the negative branch.}
\label{para}
\end{figure}

From the resonance conditions represented by the delta functions in \Eq{collop} for the absorption and emission of phonons, we obtain $\varepsilon_{p'}-\varepsilon_{p} = \hbar\omega_q >0$ and $\varepsilon_{p'}-\varepsilon_{p}=- \hbar\omega_q<0$, respectively. Consequently, two adjacent momenta $p$ and $p'$ are related by
\begin{equation}
\half \pb'^2-\half \pb^2 \pm(\pb'-\pb)=\half (\pb'-\pb)(\pb'+\pb\pm 2)=0\,,
\end{equation}
where we introduce the dimensionless momentum
\begin{align}   \label{dlessp}
    \pb&\equiv p/(mc) \,,
\end{align}
and $\hbar q/(mc)=\pb'-\pb$. Thus each $\pb$ is coupled with two other momenta $-\pb\pm 2$.
If we start with some $\p0$, then all the momenta coupled directly or indirectly
to $\p0$ can be reached by recursively applying the following formula,
\begin{align}
        \pb_{i\pm1} = -\pb_i\pm2(-1)^i\,, \quad  i=0, \pm 1, \pm 2,\ldots .
\end{align}
The solutions of this recursive formula are the mutually disjoint momentum subspaces each represented by $\p0$,
\begin{align}   \label{p0pm}
    \pb_{0;\pm i} \equiv (-1)^i (\p0 \mp 2i)\,, \quad  i=0,1,2,\ldots .
\end{align}
\Fig{para} illustrates the disjoint momentum subspaces for a few values of $\p0$. When $\p0=0$, the positive and negative branches are mirror image of each other along the $\p0=0$ axis, see \Fig{para}(i). At $\p0=\pm 1$, both branches will merge, see \Fig{para}(iii) for the case of $\p0=1$.
By varying $\p0$ across the interval $-1\leq \p0\leq1$, all the momentum states are covered. The resonance conditions therefore divide the momentum space
into infinite disjoint sets $\{\pb_{0;i}\}$, each set is represented by $\p0$ in the interval $-1\leq \p0 \leq 1$.

The existence of the disjoint momentum subspaces is a special feature of a 1D system. In higher dimensions, the energy and momentum conservation equations will give rise to, in the two dimensional case for instance, a set of closed lines. These lines may intersect with each other. As a result, all the momentum states become connected.

\section{Eigenvalue problem of collision operator}
\label{sec-eigcoll}

\subsection{Biorthogonal basis}

By specializing to a particular momentum subspace $\{\pb_{0;i}\}$ connected to $\p0$, let us consider the eigenvalue problem
\begin{align}   \label{Kphi}
    \Khat \phi_\mu(\pb_{0;\pm i})= - \lam_\mu(\p0) \phi_\mu(\pb_{0;\pm i}) \,,
\end{align}
where we anticipate the fact that the eigenvalue is a function of $\p0$.
We will label the components of the eigenvectors as
\begin{align}   \label{phipm}
    \phi^\pm_{\mu;i}(\p0) \equiv \phi_\mu(\pb_{0;\pm i})   \,, \quad i=0,1,2, \ldots.
\end{align}
When $\p0$ lies in the interval $(-1,1)$, the eigenvalue problem takes the form
\begin{subequations}
\begin{align}
 \label{eprob0}
    -\lb_\mu \phi_{\mu;0} &= [1+n(2+2\pb_0)]\phim_{\mu;1}\no
    &\quad  -[n(2-2\pb_0)
     -n(2+2\pb_0)]\phi_{\mu;0} \no
     &\quad+[1+n(2-2\pb_0)]\phip_{\mu;1} \,, \qquad  |\p0|<1\,,\\
    -\lb_\mu \phi^\pm_{\mu;i} &= n(2|\pb_{0;\pm i}|-2) \phi^\pm_{\mu;i-1} \no
     &\quad-[1+n(2|\pb_{0;\pm i}|-2)+n(2|\pb_{0;\pm i}|+2)] \phi^\pm_{\mu;i} \no
     &\quad+[1+n(2|\pb_{0;\pm i}|+2)] \phi^\pm_{\mu;i+1} \,, \no
    & \qquad\qquad\qquad 1<|\pb_{0;\pm i}|,\quad i=1,2,3,\ldots,
   \label{eprobn}
\end{align}
\end{subequations}
where we define the dimensionless eigenvalue
\begin{align}   \label{lambar}
   \lb_\mu
   \equiv
     \frac{\rho_M  \hbar^2 c}{m\Delta_0^2} \lam_\mu \,,
\end{align}
and omit the $\p0$ dependence on $\phi^\pm_{\mu;i}$ to simplify the notation. The corresponding dimensionless time is then $\tb\equiv m\Delta_0^2/(\rho_M  \hbar^2 c) t$.

For the boundary value $\p0=\pm1$, \Eq{eprob0} will be modified. In this case, there is overlap in the momentum states, $\pb^+_{0;i\pm 1}=\pb^-_{0;i}$, see \Fig{para}(iii). As a result, $\phi^+_{\mu;i\pm 1}=\phi^-_{\mu;i}$. Therefore, the components of the eigenvectors that need to be considered reduce by almost half compared to the case $\p0\neq \pm 1$. At this point, we make use of the freedom to choose, as independent components of the eigenvectors, elements of the negative branch $\phi^-_\mu$ when $\p0=1$, and elements of the positive branch $\phi^+_\mu$ when $\p0=-1$. We then have
\begin{align}   \label{ep10} \tag{\ref{eprob0}$'$}
     -\lb_\mu \phi_{\mu;0} &=-n(4)\phi_{\mu;0}+ [1+n(4)]\phi^\mp_{\mu;1}  \,, & \p0=\pm1 \,,
\end{align}
whereas the rest of $\phi^\pm_{\mu;i}$ are still related to $\phi_{\mu;0}$ via \Eq{eprobn}.

The eigenvalue problem can now be compactly written as a matrix equation,
\begin{align}   \label{eprob}
    \K(\p0)\cdot \phiv_\mu(\p0)
    =
    -\lb_\mu(\p0) \phiv_\mu(\p0) \,.
\end{align}
The eigenvector $\phiv_\mu$ is a column matrix,
\begin{align}   \label{phiv}
    \phiv_\mu(\p0) \equiv\left( \begin{array}{c}
              \vdots  \\
              \phim_{\mu;i}(\p0) \\
              \vdots   \\
               \phim_{\mu;1}(\p0)  \\
               \phi_{\mu;0}(\p0) \\
               \phip_{\mu;1}(\p0)\\
               \vdots \\
               \phip_{\mu;i}(\p0) \\
              \vdots  \end{array}   \right) \,,
\end{align}
whereas the collision operator $\K$ is a tridiagonal matrix, with components $K_{ij}$ that can be read directly from \Eqs{eprob0} or \eqref{ep10}, and \eqref{eprobn}. They are
\begin{subequations}
\begin{align}   \label{Kcomp}
    K_{0,0}&= -n(2-2\p0)-n(2+2\p0)\,, &\quad |\p0| &<1\,,\\
    K_{0,\pm 1}&=1+n(2 \mp \p0) \,, &\quad |\p0| &<1\,,\label{Kcompb}
\end{align}
and
\begin{align}
    &K_{\pm i,\pm i}= -1-n(2 |\pb_{0;\pm i}|-2)-n(2 |\pb_{0;\pm i}| +2) \,,\\
    &K_{\pm i,\pm(i+ 1)} = 1+n(2|\pb_{0;\pm i}|+ 2) \,.\\
    &K_{\pm i,\pm(i- 1)} = n(2|\pb_{0;\pm i}|- 2) \,,
\end{align}
\end{subequations}
For $\p0=\pm 1$, in place of \Eqs{Kcomp} and \eqref{Kcompb}, we have
\begin{align}
    K_{0,0}&= -n(4)\,, &\qquad \p0&=\pm 1\,, \tag{\ref{Kcomp}$'$} \\
    K_{0,\pm 1}&=1+ n(4) \,, &\qquad \p0&=\pm 1\,. \tag{\ref{Kcompb}$'$}
\end{align}

The canonical equilibrium distribution function $\exp(-\btb \pb_{0;\pm i}^2/2)$ is the steady state of the collision operator with zero eigenvalue $\lam_0=0$, where $\btb=1/\Tb$ and $\Tb$ is the dimensionless temperature
\begin{align}   \label{dlessT}
    \Tb&\equiv k_BT/(mc^2) \,,
\end{align}
This can be inferred from \Eqs{lam+} and \eqref{lam+P01} in the Appendix \ref{sec-positive}, with the help of relation \Eq{leftR} below. We denote the components of the steady state vector by
\begin{align}   \label{vphiv}
    \varphi^\pm_i(\p0)\equiv \frac{e^{-\btb \pb_{0;\pm i}^2 /2}}{Z(\p0)}    \,, \quad i=0,1,2,\ldots\,.
\end{align}
With the normalization factor
\begin{align} \label{Z}
Z(\p0)\equiv \sum_{i=-\infty}^\infty e^{-\btb \pb_{0; i}^2 /2}\,,
\end{align}
we have $\vphiv=\phiv_0$.

Due to the non-hermiticity nature of $\K$, we also need to consider the left eigenvalue problem $\chiv^\dg_\mu\cdot \K=-\lb_\mu \chiv^\dg_\mu$ \cite{Morse}, or equivalently,
\begin{align} \label{LEprob}
        \K^\dg(\p0)\cdot\chiv_\mu(\p0)
        =
        -\lb_\mu(\p0) \chiv_\mu(\p0) \,,
\end{align}
where $\chiv_\mu$ is the left eigenvector of $\K$. It can be shown that the components of $\K$ satisfy the relation (no summation over $i$ and $j$)
\begin{align}   \label{Kadj=K}
  K^\dg_{ij} =\varphi^{-1}_i K_{ij} \varphi_{j}\,.
\end{align}
In matrix form, we write this as
\begin{align}   \label{Kadj=K*}
  \K^\dg \equiv \vphiv^{-1} \circ \K \circ \vphiv \,,
\end{align}
where the $\circ$-product is defined by the right hand side (RHS) of \Eq{Kadj=K}.
It follows from \Eq{eprob} that
\begin{align}   \label{innleftb}
  \K^\dg \cdot (\vphiv^{-1}\circ \phiv_\mu(\p0))
  =
  \lb_\mu(\p0) \vphiv^{-1}\circ \phiv_\mu(\p0)\,.
\end{align}
A comparison with \Eq{LEprob} shows that the left eigenvectors are related to their right counterparts by
\begin{align}   \label{leftR}
    \chiv_\mu = \vphiv^{-1} \circ \phiv_\mu \,,
\end{align}
or equivalently,
\begin{align}   \label{leftRb}
      \phiv_\mu = \vphiv \circ \chiv_\mu \,.
\end{align}

The left and right eigenvectors are biorthogonal \cite{Morse}. This means that they are orthogonal and complete in the sense of
\begin{align}   \label{orthon}
   \chiv^\dg_\mu(\p0) \cdot \phiv_\nu(\p0) = \del_{\mu\nu} \,,\\
  \label{comp}
    \sum_\mu \phiv_\mu(\p0) \cdot  \chiv^\dg_\mu(\p0) = \textbf{I} \,,
\end{align}
respectively, where $\textbf{I}$ is an infinite dimensional identity matrix.
It should be noted that \Eqs{orthon} and \eqref{comp} apply to each $\p0$ subspace, and there is a separate complete set of orthonormal eigenvectors for each $\p0$ subspace.

In this space, we define the inner product of two vectors $\gb$ and $\fb$ as
\begin{align}   \label{iprod}
    \l \gb(\p0) |\fb(\p0)\r &\equiv [\vphiv^{-1}\circ \gb(\p0)]^\dg  \cdot \fb(\p0) \no
    &= \sum_{i=-\infty}^\infty \varphi^{-1}_i g^*_{i}(\p0) f_{i}(\p0) \,.
\end{align}
This inner product should be compared to the usual inner product in continuous space,
\begin{align}   \label{uprod}
    \int_{-\infty}^\infty d\pb \,\, \varphi^{-1}_\text{eq}(\pb) g^*(\pb) f(\pb) \,.
\end{align}
\Eq{orthon} is then equivalent to the norm of the right eigenvectors, $\l \phiv_\mu | \phiv_\nu\r = \del_{\mu\nu}$. Using the orthogonal relation \eqref{orthon}, we show in Appendix \ref{sec-positive} that the eigenvalues are non-negative.

It is a common feature in dissipative systems that the eigenvectors belong to the generalized space\cite{gensp}, i.e., they are not objects in the Hilbert space. In other words, the pair of vectors $\{ \phiv_\mu,\chiv_\mu\}$ are generalized vectors. The vectors $\phiv_\mu$ belong to the space of test function and is normalizable. On the other hand, $\chiv_\mu$ belong to the dual space of the test function and are not normalizable. For example, consider the steady mode that has components $\chi_{0;i}=1$ for all $i$. $\chiv_0$ is therefore non-normalizable under the inner product \eqref{iprod}.

To some extent, the non-hermiticity of the operator $\K$ is artificial, since one may transform it to a hermitian operator through a similarity transformation. A symmetrized version of $\K$ as introduced in Ref.~\cite{Tanaka} achieves this aim. The transformed left and right eigenvectors in this case become normalizable.

\subsection{Time evolution}

We can determine the time evolution of an initially continuous distribution function $f(\pb)$ by first discretizing it into a set of column vectors $\fb(\p0)$ (see \Eq{phiv} for an explicit form of the vector), for all $ \p0$. The components of the vector $\fb(\p0)$ are labeled by
\begin{align}   \label{vPhip0comp}
    f^\pm_i(\p0) &\equiv   f(\pb_{0;\pm i})    \,, & i&=0,1,2,\cdots\,.
\end{align}

Using the orthonormality of the basis vector \eqref{orthon}, we can expand $\fb(\p0)$ in terms of $\phiv_\mu(\p0)$. The time evolution of $\fb(\p0)$ can then be worked out,
\begin{align}   \label{Phiexp}
    \fb(\p0;\tb) &= e^{\K \tb}\cdot  \fb(\p0;0) \no
      &=\sum_\mu c_\mu(\p0) e^{-\lb_\mu(\p0) \tb} \phiv_\mu(\p0)\,,
\end{align}
where the expansion coefficient is
\begin{align}   \label{cmu}
    c_\mu(\p0) &= \chiv^\dg_\mu(\p0) \cdot \fb(\p0;0) \,.
\end{align}

The time evolution of the inner product is
\begin{align}   \label{iprodt}
    \l \fb(\p0;\tb) |\fb(\p0;\tb)\r
    &= \sum_{\mu} |c_\mu(\p0)|^2  e^{-2\lb_\mu(\p0)\tb}  \,,
\end{align}
after using \Eqs{leftR} and \eqref{orthon}. It is a monotonously decaying function, in accordance to the Markovian nature of the kinetic equation under the $\lam^2t$-approximation.

Since the collision operator can be written in terms of a linear combination of differential operators, and the eigenvalues $\lb(\pb)$ of adjacent momentum states vary continuously, it is intuitive that a distribution function that is initially continuous to maintain its continuity as it evolves, just like the time evolution governed by a finite order differential operator in the usual case. This is supported by numerical simulation on the time evolution of distribution function.

\section{Continued fraction method}
\label{sect-eigenvalue}

The tridiagonal nature of the collision operator $\K$ prompts us to solve the eigenvalue problem by the continued fraction method \cite{Morse,Risken}. Hereafter, we omit the eigenvector index $\mu$ for simplicity of notation. We introduce the ratio
\begin{align}   \label{Ft}
    F^\pm_i(\p0) &\equiv  \frac{\phi^\pm_{i-1}(\p0)}{\phi^\pm_i(\p0)}\,,& i&=1,2,3,\ldots\,,
\end{align}
which has the advantage of reducing the independent components in the eigenvalue equation. In terms of $\phi^\pm_i$, we need the values of two components of $\phi^\pm_i$, to determine the values of rest of the other components. But in terms of $F^\pm_i$, the value of one component will fix the values of the other components.

We begin by rewriting \Eqs{eprob0} and \eqref{eprobn} in terms of $ F^\pm_i$. Since we restrict our consideration to a specific $\p0$ momentum subspaces, we omit the $\p0$ dependence on $F^\pm_i$ for simplicity. In what follows, we will regard $F^\pm_i$ as a function of $\lb$. We get
\begin{subequations}
\begin{align}   \label{Fteq3}
    F^+_1(\lb)&= \frac{1+n(2-2 \p0)}{-\lb +n(2-2\p0)+n(2+2 \p0)-\cfrac{1+n(2+2\p0)}{F^-_1(\lb)}} \,,\no
            &\qquad\qquad\qquad\qquad\qquad\qquad\qquad\qquad |\p0|<1\,,\\
            \label{Fteq2}
    F^\pm_i(\lb)& =x^\pm_i-\lb r^\pm_i+\frac{y^\pm_i}{F^\pm_{i+1}(\lb)}\,, \qquad  i=1,2,3,\ldots,
\end{align}
\end{subequations}
whereas \Eq{ep10} becomes
\begin{align}   \label{Ftp01} \tag{\ref{Fteq3}$'$}
    F^\mp_1(\lb) &= \frac{1+n(4)}{-\lb+n(4)} \,, &\quad \p0&= \pm 1\,.
\end{align}
The coefficients are
\begin{subequations}
\begin{align}   \label{xry}
    x^\pm_i&= 1+\frac{1+n(2|\pb_{0;\pm i}|+2)}{n(2|\pb_{0;\pm i}|-2)} \,,\\
    r^\pm_i&= \frac{1}{n(2|\pb_{0;\pm i}|-2)} \,,\\
    y^\pm_i&=1-x^\pm_i=-\frac{1+n(2|\pb_{0;\pm i}|+2)}{n(2|\pb_{0;\pm i}|-2)} \,.
\end{align}
\end{subequations}

When we iterate $F^\pm_1$ using \Eq{Fteq2}, we get continued fraction. In practice, we truncate the iterations at a large enough value of $i=N\gg 1$. This approximation reduces the dimensionality of the vector space to $2N+1$ for $\p0\neq 1$, and $N+1$ for $\p0=\pm1$. In this way, the $N$-th approximant of $F^\pm_1$ is
\begin{align} \label{Nite}
       F^\pm_1(\lb)
       &=x^\pm_1-\lb r^\pm_1+
       \frac{y^\pm_1}{x^\pm_2-\lb r^\pm_2+
       \cfrac{y^\pm_2}{\ldots+
       \cfrac{\vdots}{\ldots+
       \cfrac{y^\pm_{N-1}}{ F^\pm_N(\lb)
       }}}} \,.
\end{align}
In the limit of large $N \gg 1$, we find that
\begin{align}   \label{xy}
    x_N^\pm,\, r_N^\pm &\to e^{2\btb |\pb_{0;\pm N}|}\,, \\
    y_N^\pm &\to -e^{2\btb |\pb_{0;\pm N}|} \,. \label{r}
\end{align}
Consequently, \Eq{Fteq2} gives us the boundary value,
\begin{align}   \label{Ftinfty}
    F^\pm_N(\lb)&\to ( 1-\lb) \, e^{2\btb |\pb_{0;\pm N}| }\,.
\end{align}
The continued fraction \eqref{Nite} with the boundary value \eqref{Ftinfty} defines $F_1^\pm$ as a function of $\lb$. By equating the RHS of \Eqs{Fteq3} or \eqref{Ftp01} to $F^+_1$ of \Eq{Nite}, we can determine the eigenvalues, which are the intersections of these functions of $\lb$.
The components of the eigenvector can then be obtained by iteration,
\begin{align}   \label{phiN}
    \phi_N = \frac{1}{F_N}  \phi_{N-1}&= \frac{1}{F_N} \frac{1}{F_{N-1}}  \phi_{N-2} \no             &= \cdots
            = \left( \prod_{i=1}^N \frac{1}{F_i}  \right) \phi_0\,.
\end{align}

This method had been used to solve differential kinetic equations \cite{Risken} and Schr\"odinger equation \cite{CFSE} too, where the continuous nature of the system permits the eigenfunction expansion in terms of a complete set of orthogonal functions, leading to a tridiagonal structure of the equations in terms of the coefficients of expansion. On the other hand, due to the discrete nature of the system we consider here, we cannot expand the eigenvector in terms of continuous orthogonal functions. We instead apply the continuous fraction method directly to the components of the eigenvector at discrete set of momentum. This method is particularly suitable in solving difference kinetic equations with discretely coupled momentum or position states.

\section{Spectrum}
\label{sec-spectrum}

\begin{figure}[htb]
  \centering
\includegraphics[width=3.2in]{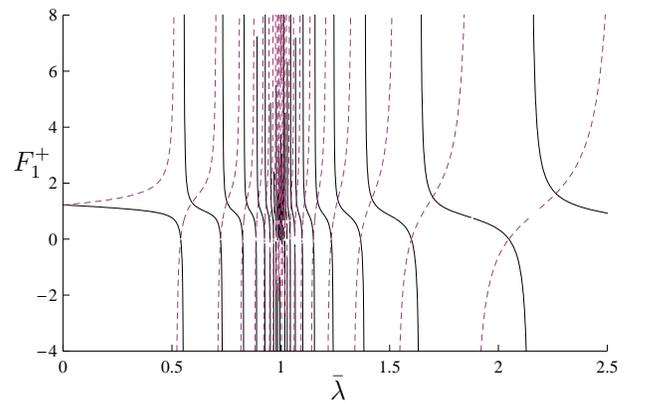}
\caption{For $N=50$, $ \Tb=10$ and $ \p0=0$. Solid curves are $F^+_1(\lb)$. Dotted curves are the RHS of \Eq{Fteq3}. The intersections of both curves give the eigenvalues.}
\label{F1pmN50}
\end{figure}

Due to a reflection symmetry of the collision operator under $\p0\leftrightarrow-\p0$, we have $\lb_\mu(\p0)=\lb_\mu(-\p0)$, see Appendix \ref{sec-sym}. Therefore, we consider only the eigenvalues for $\p0 \geq 0$. \Fig{F1pmN50} illustrates how the eigenvalues are determined. The solid curves are $F^+_1(\lb)$ whereas the dotted curves are the RHS of \Eq{Fteq3}. The intersections of both curves give the eigenvalues of the collision operator.

\subsection{Accumulation point}
\label{sec-accpoint}

\begin{figure}[htb]
    \centering
\includegraphics[width=3.2in]{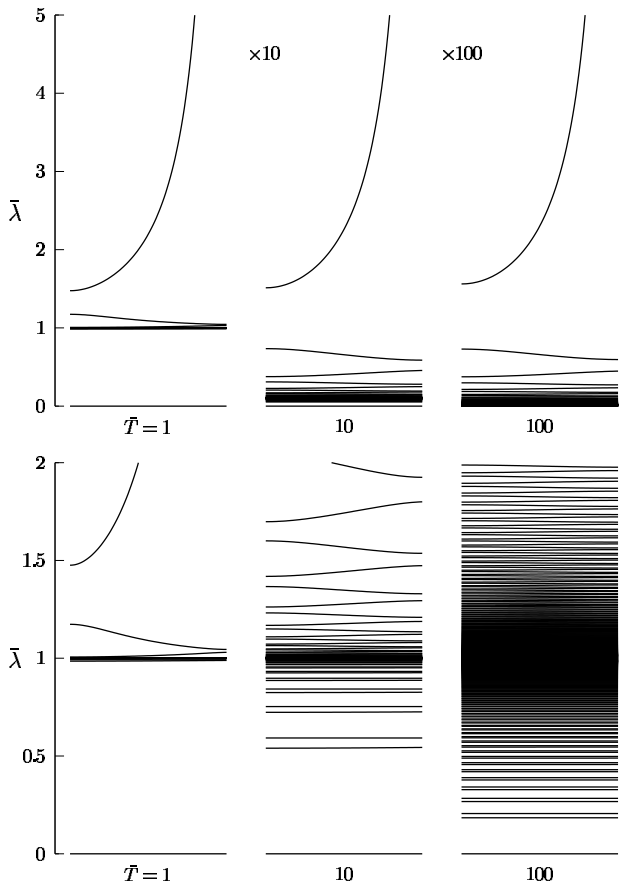}
\caption{Spectrum of the eigenvalues as $\p0$ varies across $0\leq \p0\leq 1$ for three different temperatures, $\Tb=1, 10,100$. The top figure is plotted with different vertical scales. The bottom figure is a magnified view of the top figure for the interval $0\leq \lb \leq2$.}
\label{spec}
\end{figure}

\Fig{spec} shows that the spectrum is discrete and bounded from below by the steady mode $\lb_0=0$. Furthermore, it is bounded from above by a largest value labeled by $\lb_{-1}$ below.
The most striking feature of the spectrum is the existence of a common accumulation point at $\lb=1$, which is the special feature in the solution of the difference collision operator.

To see how the accumulation around $\lb=1$ occurs, let us consider the matrix $\K$. As mentioned earlier, it has a tridiagonal structure according to \Eqs{eprob0} and \eqref{eprobn}. For large momentum, or equivalently large $i\gg 1$, \Eq{eprobn} approaches
\begin{align}   \label{eprobnlargeP}
     -\lb \phi^\pm_{i} &= - \phi^\pm_{i} + \phi^\pm_{i+1} \,, \qquad i \gg 1.
\end{align}
The portion of the characteristic equation of $\K$ involving the large $i\gg 1$ elements of the positive branch $\phip$
reduces to  (recall the difference of the sign in $\lb$ from $\lam$ in \eqref{lambar})
\begin{align}   \label{cheq}
    |\K+\lb {\bf I}| = \left| \begin{array}{ccccc}
              \ddots & & \vdots & \\
              & \lb-1 & 1 & 0 & \\
              \cdots & 0 & \lb-1 & 1 & \cdots \\
             & 0 & 0 & \lb-1 & \\
             & & \vdots & & \ddots   \end{array}   \right| =0\,,
\end{align}
which gives rise to degenerate value
at $\lb$=1. The same observation occurs for the negative branch $\phi^-$ with $i\ll -1$. Therefore, the eigenvalues accumulate around $\lb =1$.

We recall that we have approximated the continued fraction by truncating it at a large momentum $\pb_{0;\pm N}$. As we increase $N$, new eigenvalues emerge increasingly close to the accumulation point, and are distributed evenly on both sides of it. On the other hand, the values of the smaller and larger values
away from $\lb= 1$ are not affected much by the increase of $N$, as long as large enough iterations $N$ had been carried out.

In view of the existence of accumulation point and a maximum value of $\lb$,
we label the values in the following way. For $\lb<1$, we label them with increasing positive integer according to their magnitude, i.e., $0=\lb_0<\lb_1<\lb_2< \ldots <1$. For $\lb>1$, we label the values
with negative integer in decreasing order of magnitude, with the largest value of $\lb$
labeled by $\lb_{-1}$, i.e., $ 1< \ldots <\lb_{-2} <\lb_{-1} < \infty$.

\subsection{Band and pairing structure}

Due to the disjointness in the momentum subspaces, the eigenvalues in general depend on the representative momentum $\p0$ and vary continuously with it, except the steady mode which remains at zero.  Consequently, the spectrum exhibits a band structure as we vary $\p0$, which is more prominent for larger values of $\lb$. In each momentum subspace, the spectrum is bounded by the largest
value  $\lb_{-1}$ (i.e., the smallest value of $\lam$),
which approaches infinity as $|\p0| \to 1$. For the boundary values $\p0=\pm 1$, we show in Section \ref{sec-disappear} below that all values
with odd indices $\lb_{\pm 1}, \lb_{\pm 3}, \ldots$, drop out from the spectrum. $\lb_{-2}$ then becomes the maximum value
in this case. The existence of a maximum value (i.e., the minimum eigenvalue)
in this case is distinctively different from the spectrum of a differential operator that is usually unbounded from above.

The bottom plot of \Fig{spec} also suggests a pairing of the values for $\lb<1$. This observation can be explained as follows. As a function of $\lb$, $F^+_1$ \eqref{Nite} has infinite numbers of poles
in the real axis of $\lb$.
Whenever the variable $\lb$ crosses a pole, the RHS of \Eq{Fteq3} (dotted curves in \Fig{F1pmN50}) will intersect $F^+_1$ (solid curves in \Fig{F1pmN50}) at two eigenvalues situated close to the pole on both sides. This gives rise to the pairing structure that is more obvious for $\lb<1$.

\subsection{Temperature dependence}

In the low temperature limit $\Tb \to 0$, \Eqs{eprob0} and \eqref{eprobn} go into
\begin{subequations}
\begin{align}   \label{eprobLowT}
    - \lb \phi_0 &=  \phi^-_1 + \phi^+_1 \,,\\
     - \lb \phi^\pm_{i} &= - \phi^\pm_{i} + \phi^\pm_{i+1} \,.
\end{align}
\end{subequations}
In this limit, the characteristic equation of $\K$ takes the form
\begin{align}   \label{cheqLowT}
    |\K+\lb {\bf I}| = \left| \begin{array}{ccccccc}
              \ddots & & & \vdots & \\
              & 1& \lb-1 & 0 & 0 & 0 &\\
              \cdots & 0& 1 & \lb & 1 & 0 & \cdots \\
             & 0 & 0 & 0& \lb-1 & 1 & \\
             & & &\vdots & & &\ddots   \end{array}   \right| =0 \,.
\end{align}
Therefore, there is one eigenvalue at $\lb=0$, whereas the eigenvalue
corresponding to the accumulation point $\lb=1$ becomes infinitely degenerate.

As the temperature increases, the eigenvalues are distributed evenly away from the accumulation point. As a result, the larger and the smaller eigenvalues are affected in an opposite manner by the change in temperature. When temperature increases, the magnitude of the larger
values of $\lb$ increases whereas the smaller ones decreases.

\section{Eigenfunctions}
\label{sec-approx}

In the usual eigenvalue problem of a finite order differential operator, we consider an equation of the form
\begin{align}   \label{EigvUsual}
   \Khat' \phi'_\mu(\pb)=-\lam'_\mu \phi'_\mu(\pb)\,,
\end{align}
where the eigenvalues are independent of $\pb$. However,
In the 1D model we consider in the present case, we already learned that because of the disjoint nature of the momentum space, the eigenvalues are generally $\pb$-dependent, except for the degenerate steady mode. Consequently, we consider an equation of the form
\begin{align}   \label{Eigvhere}
   \Khat \phi_\mu(\pb)=-\lam_\mu(\pb) \phi_\mu(\pb)\,,
\end{align}
where the eigenvalues now depend on $\pb$. As a result, the completeness and orthogonality relations only hold for components of $\phi(\pb)$ belonging to the same momentum subspaces $\phi(\pb_{0;\pm i})$. Hence, in general the idea of an eigenfunction with $\pb$ as a continuous variable does not exist in the exact sense of \Eq{EigvUsual}. However, when $\lb_\mu(\p0)$ are approximately degenerate, for example for $\lb_\mu<1$, or for $\lb_\mu>1$ at high temperature, see \Fig{spec}, one can construct approximate eigenfunctions of the collision operator $\Khat $ that is at least piecewise continuous.

In this section, we regard the eigenvalues and eigenvectors as continuous functions of $\p0$ in the range $[-1,1]$. In the limit $\p0 \to \pm1$, the left side
limit  ($\pm 1^-$) and the right side limit ($\pm 1^+$)
of the component of the eigenvector at odd value of momentum, $\pb_\text{odd}\equiv \pm1, \pm3,\pm5,\ldots$, may not be the same. This will be shown to cause the disappearance of the eigenvalue at $\p0=\pm1$ for odd $\mu$. After clarifying this fact, we construct the piecewise continuous functions and compare them to the approximate eigenfunctions obtained by truncating the collision operator to become a second order differential operator at the high temperature and large momentum limit derived in Ref.~\cite{Tanaka}.

\subsection{Disappearance of eigenvalues with odd indices when $\p0=\pm1$}
\label{sec-disappear}

An interesting feature of the spectrum occurs in the boundary case $|\p0|= 1$. In this case, both branches of the momentum $\pb_{0;\pm i}$ coincide, thus reducing the independent components of the eigenvectors by almost half, see \Fig{para}(iii). It is found that the eigenvalues with odd indices in the $|\p0|< 1$ spectrum, i.e., $\lb_{\pm1},\lb_{\pm3},\ldots$, disappear from the spectrum of $|\p0|= 1$. As a result, there is a discontinuity of $\phi_\mu$ at $\pb_\text{odd}$ for these eigenvalues.

To understand the disappearance, let us consider the limit $\p0\to 1$
 (the other limit $\p0\to -1$ can be worked out similarly).
 In \Eq{eprob0}, the factor $n(2-2\p0)$ diverges in the limit $\p0\to 1$. In order to avoid this divergence, and to have \Eq{eprob0} goes into \Eq{ep10} in the same limit, we require
\begin{align}   \label{P1-}
    \underset{\p0\to 1^-}{\lim} \left\{ n(2-2 \p0) \phi_{\mu;0} -[1+n(2-2\p0)] \phip_{\mu;1} \right\} =0\,.
\end{align}
where $\p0\to 1^-$ or $ 1^+$ means that $\p0$ approaches 1 infinitesimally from the left or right side, respectively.
Since $1+n(2-2\p0)=n(2-2\p0) \exp(-|2-2\p0|\btb )$, we conclude that
\begin{align}   \label{P1-phi}
    \underset{\p0\to 1^-}{\lim} \phi_{\mu;0}=  \underset{\p0\to 1^-}{\lim} \phip_{\mu;1} \,,
\end{align}
From the notation defined in \Eq{phipm}, \Eq{P1-phi} is equivalent to
\begin{align}   \label{P1-P0}
    \underset{\p0\to 1^-}{\lim} \phi_\mu(\p0)&=  \underset{\p0\to 1^-}{\lim} \phi_\mu(-\p0+2) \no
    &=  \underset{\p0\to 1^+}{\lim} \phi_\mu(\p0)  \,.
\end{align}
If we consider the limit $\p0 \to -1^-$, we will instead obtain
\begin{align}   \label{P1-P02}
\underset{\p0\to -1^-}{\lim} \phi_\mu(\p0)=  \underset{\p0\to -1^-}{\lim} \phi_\mu(-\p0-2) =  \underset{\p0\to -1^+}{\lim} \phi_\mu(\p0)\,.
\end{align}

Through a numerical study, we find that the requirements \eqref{P1-P0} and \eqref{P1-P02} are satisfied by solutions that correspond to eigenvalues with even indices, $\lb_0, \lb_{\pm 2}, \lb_{\pm 4}, \ldots$. These eigenvalues are always finite. As for the solutions that correspond to eigenvalues with odd indices, $\lb_{1}, \lb_{\pm 3}, \ldots$, the limit happens to vanish and both requirements are satisfied. \Eq{eprobn} then implies that all other components of the eigenvectors, $\phi^\pm_{\mu;i}$, also vanish, and this set of eigenvalues drops out from the spectrum of $\p0=\pm 1$. The maximum
 value
 is then $\lb_{-2}$, which is always finite.

An exceptional case occurs for $\mu=-1$. The limit $\p0\to 1^-$ but $\p0 \neq 1$ in the components $\phi_{\mu;0}$ and $\phi^+_{\mu;1}$ is non-vanishing, and \Eq{P1-phi} is not satisfied. Hence the largest
 value
 $\lb_{-1}$ becomes increasing large as $\p0$ approaches $1$ from the left side, due to the factor $n(2-2\p0)$.
An infinite decay rate means that this mode vanishes almost immediately in the time evolution. It is then the second largest
 value
 $\lb_{-2}$, which is always finite, that is observed in the time evolution of the state. At exactly $\p0=1$, to avoid the divergence of $n(2-2\p0)$, the components $\phi_{\mu;0}$ and $\phi^+_{\mu;1}$ have to vanish identically. Therefore, $\lb_{-1}$ drops out from the spectrum at $\p0=\pm 1$.

Now we show that for $\mu=1, \pm 3, \pm 5, \ldots$, $\phi_\mu(\pb)$ has discontinuities in the neighborhood of $\pb_\text{odd}\equiv\pm 1, \pm 3, \pm 5, \ldots$. Using \Eqs{eprob0} and \eqref{eprobn}, we consider the sum
\begin{align}   \label{sumpp}
   & \lb_\mu(\phi_{\mu;0}+\phip_{\mu;1}) = n(2+2\p0)\phi_{\mu;0}+n(6-2\p0)\phip_{\mu;1} \no
    &\quad -[1+n(2+2\p0)]\phim_{\mu;1}-[1+n(6-2\p0)] \phip_{\mu;2} \,.
\end{align}
Since the limit on the both sides of \Eq{P1-phi} happens to vanish as already shown in the previous paragraph, \Eq{sumpp} implies that
\begin{align}   \label{sumpp2}
  \underset{\p0\to 1^-}{\lim} ( \phim_{\mu;1}+\phip_{\mu;2})
  &=\underset{\p0\to 1^-}{\lim} [\phi_\mu(-2-\p0)+\phi_\mu(\p0-4)]\no
 &  =\underset{\p0\to -3^-}{\lim}\phi_\mu(\pb)+\underset{\p0\to -3^+}{\lim}   \phi_\mu(\pb)=0 \,.
\end{align}
For $\mu=-1$, the limit on both sides of \Eq{P1-phi} do not vanish, but they are of equal magnitude but of different sign. Hence in the limit $\p0\to 1^-$ \Eq{sumpp} again leads to \Eq{sumpp2}. For other $\pb_\text{odd}$, we can then derive inductively that
\begin{align}   \label{lim1}
  \underset{\p0\to \pb_\text{odd}^-}{\lim} \phi_\mu(\pb)=- \underset{\p0\to \pb_\text{odd}^+}{\lim}   \phi_\mu(\pb)
\end{align}
for all odd $\mu$.
In other words, there is a discontinuity of $\phi_\mu(\pb_\text{odd})$ at $\pb_\text{odd}$ for the eigenvalues with odd indices, whereas $\phi_\mu(\pb_\text{odd})=0$. This can be seen from the (ii) and (iv) plots in \Fig{pseudo}.

\subsection{Piecewise continuous functions}
\label{sect-eigev}

\begin{figure}[tb]
   \centering
\includegraphics[width=3.4in]{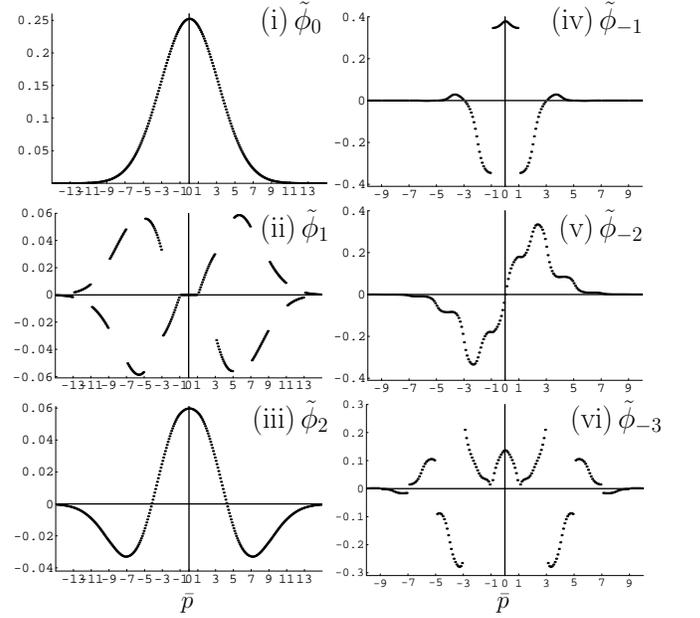}
\caption{Approximate eigenfunctions of the collision operator, only the steady mode (i) is an exact eigenfunction. Each plot is made by a collection of points for momentum subspaces connected to $\p0=0,\pm 0.1,\pm 0.2,\ldots ,\pm1$, where $\Tb=10$.}
\label{pseudo}
\end{figure}

The piecewise continuous functions for each eigenvalue indices can be constructed as follows.
After normalizing $\phiv_{\mu;0}= \phiv_\mu(\p0)$ according to $\l\phi_\mu(\p0)|\phi_\mu(\p0)\r=1$, we are still left with the freedom to choose the sign of the representative component $\phi_\mu(\p0)$. Once this sign is fixed, the signs for the rest of $\phi_\mu(\pb_{0;\pm i})$ are automatically fixed. We use this freedom to set $ \phi_\mu(\p0) \geq 0$ for $\p0 \geq 0$. When $\p0<0$, the reflection symmetry \eqref{phixphivcomp} discussed in Appendix \ref{sec-sym} gives rise to even or odd functions, $ \phi_\mu(\p0)=\phi_\mu(-\p0)$ or $-\phi_\mu(-\p0)$, respectively, depending on the eigenvalue index $\mu$.

In this way, we obtain even functions $\phit_\mu(\pb)$ whenever $\mu=0, 2, 4, \ldots,  -5, -3, -1$, and odd functions whenever $\mu=1, 3, 5, \ldots, -6, -4, -2$. It happens that the functions $\phit_\mu(\pb)$ for $\mu=\pm 1, \pm 3, \pm 5, \ldots$, can at best be made piecewise continuous, due to the discontinuities at $\pb_\text{odd}$, which is a consequence of the disappearance of the eigenvalues from the spectrum for $|\p0|= 1$, as discussed in Section \ref{sec-disappear}.
In \Figs{pseudo}, we give the examples of a few $\phit_\mu$ constructed in this way. Symmetric $\phit_\mu$ occurs for $\mu=\ldots,-3,-1,0,2,4,\ldots, $ and anti-symmetric $\phit_\mu$ occurs for $\mu=\dots, -4,-2,1,3,\dots$.

As discussed in the
first part of  Section \ref{sec-approx},
$\phit_\mu(\pb)$ become the eigenfunction of the collision operator when the eigenvalues for different momentum subspaces become approximately degenerate. As can be seen from \Fig{spec}, the eigenvalues are always close to degenerate for $\lb_\mu<1$, i.e., when $\mu=0, 1, 2, \ldots$. As for $\lb_\mu>1$, the eigenvalues are approximately degenerate only in the high temperature limit for $\mu=-2, -3, -4, \ldots$.

\subsection{Comparison with solution of approximate $\Khat$}
\label{sec-highT}

In this section we will fist quote the results of an approximate collision operator obtained in Ref.~\cite{Tanaka}. The approximate collision operator is obtained by truncating the series of differential operators of $\Khat $ at the second order, supplemented by the conditions $\Tb\gg 1$ and $|\pb| \gg 1$, while maintaining $ \Tb \gg |\pb| $. Labeling the eigenfunctions and eigenvalues as $\underline{\phi}_\nu$ and $\underline{\lb}_\nu$, respectively, the approximation results in the following second order differential equation
\begin{align}   \label{KETP}
    -\frac{\Tb}{\pb} \frac{d^2 \underline{\chi}_\nu}{d\pb^2} + \frac{d \underline{\chi}_\nu}{d\pb}  =\frac{\underline{\lb}_\nu}{2} \underline{\chi}_\nu \,,
\end{align}
where the function $\underline{\chi}_\nu\equiv \varphi_\text{eq}^{-1} \underline{\phi}_\nu$, in which $\varphi_\text{eq}=\phi_0=\exp(- \pb^2/2\Tb)/\sqrt{2 \pi \Tb}$ is the steady state solution. The normalized eigenfunction is
\begin{align}   \label{normTanaka}
    \underline{\phi}_\nu(\pb)&= \frac{1}{\sqrt{2^\nu \nu! 2 \pi \Tb}} \,\, e^{-\pb^2/4\Tb}\, e^{-(\pb-\underline{\lb}_\nu \Tb)^2/4\Tb} \no
    &\qquad \times H_\nu\left[\sqrt{\Tb/2} \, (\pb/ \Tb-\underline{\lb}_\nu \big)  \right] \,,
\end{align}
with eigenvalue
\begin{align}   \label{lam'}
    \underline{\lb}_\nu=2 \sqrt{\nu/\Tb} \,.
\end{align}
The eigenfunctions satisfy the orthonormality condition
\begin{align}   \label{normcont}
  \Iii d\pb \, \varphi_\text{eq}^{-1}(\pb)  \underline{\phi}^*_\mu(\pb) \underline{\phi}_\nu(\pb)=\del_{\mu,\nu} \,.
\end{align}

A comparison between the spectrum of $\underline{\lb}_\nu$ with the solutions of the continuous fraction method in Table \ref{table1} shows that $\underline{\lb}_\nu$ closely approximate the solution obtained from the continued fraction method for the representative momentum $\p0=\pm 1$ for $\Tb=100$. The comparison also shows that the approximate solution misses the eigenvalues with odd indices in the exact collision operator. The missing modes correspond to piecewise continuous functions $\phit_\text{odd}$,  which cannot be the solutions of the second order differential equation \eqref{KETP}. Hence,
their omission from the spectrum of the approximate continuous collision operator is natural. Note that in the approximate solutions, the maximum of the non-steady modes of $\underline{\phi}_\mu$ is shifted from $\pb=0$. Therefore, $\underline{\phi}_\mu$ loses the reflection symmetry exhibited by $\phit$ along the $\pb=0$ axis.
\begin{center}
\begin{table}[htb]
\begin{tabular}{ccccccc}
$\mu$ & \hspace{1cm}& $\lb_\mu (\p0=\pm 1)$ &\hspace{1cm} &$\underline{\lb}_\nu$ &\hspace{1cm} & $\nu$\\
\hline
0 &\hspace{1cm} & 0&&0&&0 \\
2 & & 0.206&&0.200&&1 \\
4 & & 0.284&&0.283&&2 \\
6 & &  0.342   && 0.346&&3 \\
8 & & 0.390 &&   0.400 &&4 \\
\end{tabular}
\caption{Comparison of eigenvalues obtained from the continuous fraction method $\lb_\mu$ for $\p0=\pm 1$ (with missing odd indices eigenvalues), with those obtained from the approximate collision operator, $\underline{\lam}_\nu$, for $\Tb=100$.}
\label{table1}
\end{table}
\end{center}

\section{Conclusion}
\label{Concl}

We consider the relaxation process of a quantum Brownian oscillator
surrounded by a thermal equilibrium phonons in
a 1D chain as a reduced dynamics of the Liouville operator. The collision operator is a non-hermitian operator with dissipative effect that is purely quantum in nature.
The collision operator for the momentum distribution function takes the form of a difference operator.
The eigenvalue problem of the collision operator gives rise to a complete set of biorthogonal basis that belongs to the generalized space.

It is a special feature of the 1D nature of the system that the momentum and energy conservation permit momentum states to be coupled only discretely. This results in infinite sets of disjoint momentum subspaces dynamically independent of one another. The discrete nature of the collision operator facilitates the solution of the eigenvalue problem by the continued fraction method. We expect that this method is applicable as well in solving  other 1D models that have discretely coupled space or momentum subspaces.

We find that the spectrum of the collision operator
is real, discrete and  non-negative, and the system monotonically relaxes towards equilibrium. The most striking feature of the spectrum for the difference collision operator is the existence of an accumulation point. It also has an minimum eigenvalue in each momentum subspace. This is in sharp contrast to the spectrum of the usual dissipative differential operators that is unbounded from above.

Except for the steady mode, the eigenvalues of the decaying modes are non-degenerate for different momentum subspaces, giving rise to a band structure in the spectrum.
This structure originates from the disjoint nature of each momentum space.

In higher dimensions, all the momentum states are connected and this structure is lost. The general features of the results found here show up in other 1D models as well, such as in the 1D quantum Lorentz gas \cite{Gonzalo10}.

Research is under way to study the time evolution of specific distribution functions, with the aim to elucidate the consequences of the accumulation point in the spectrum on the time evolution. The accumulation point plays a dominant role in the time evolution of the system at low temperature, and it might be related to the mechanism of spontaneous emission.

\acknowledgments

We thank Professors E. C. G. Sudarshan, N. Hatano, H. Hayakawa, V. Barsegov and G. Ordonez for fruitful discussions. This work was supported by the Grant-in-Aid for Scientific Research
from the Ministry of Education, Science, Sports, and
Culture of Japan. This works was partially supported by the Yukawa International Program for Quark-Hadron Sciences (YIPQS). B.A.T. thanks Professor H. Hayakawa and the Yukawa Institute for Theoretical Physics for hospitality during the Yukawa International Seminar 2009 (YKIS 2009) and the Yukawa International Molecule Workshop (2008). B.A.T. was supported by the Malaysian Ministry of Science, Technology and Innovation (MOSTI) Postdoctoral Research Scheme (STI) when part of this work was completed.

\appendix

\section{Positivity of eigenvalues }
\label{sec-positive}

Applying the orthogonality relation \eqref{orthon} to \Eqs{eprob0} or \eqref{ep10}, and \eqref{eprobn},
and then using the relations $\pb_{0;\pm 1}=-\p0 \pm 2$ and
\begin{align}   \label{|P|}
    |\pb_{0;\pm(i+1)}|&=|\pb_{0;\pm i}|+2 \,, & i&=1,2,3,\ldots \,,
\end{align}
we can show that the eigenvalues are non-negative,
\begin{align}   \label{lam+}
    \lb_\mu &= -\chiv_\mu^\dg \cdot \K \cdot \phiv_\mu \no
            &=e^{-\btb \p0^2/2} \Big[n(2-2 \p0) |\chi_{\mu;1}^+ -\chi_{\mu;0}|^2 \no
              &\qquad\qquad\qquad + n(2+2 \p0) |\chi_{\mu;1}^--\chi_{\mu;0}|^2   \Big] \no
              & \quad+ \sum_{i=1}^\infty \Big[ e^{-\btb \pb_{0;i}^2/2} n(2|\pb_{0;i}|+2) |\chi_{\mu;i+1}^+ - \chi_{\mu;i}^+|^2 \no
              &\quad\qquad + e^{-\btb \pb_{0;-i}^2/2} n(2|\pb_{0;-i}|+2) |\chi_{\mu;i+1}^- - \chi_{\mu;i}^-|^2 \Big]\no
              &\geq 0 \,, \qquad\qquad\qquad\qquad |\p0|<1\,,
\end{align}
and
\begin{align}
    \lb_\mu &= e^{-\btb \p0^2/2}  n(4) |\chi_{\mu;1}^\mp -\chi_{\mu;0}|^2   \no
              &\quad + \sum_{i=1}^\infty  e^{-\btb \pb_{0;\mp i}^2/2} n(2|\pb_{0;\mp i}|+2) |\chi_{\mu;i+1}^\mp - \chi_{\mu;i}^\mp |^2 \no
              &\geq 0 \,, \qquad\qquad\qquad\qquad \p0=\pm 1\,, \label{lam+P01}
\end{align}
since the phonon number density $n$ is positive. Therefore, the system monotonously evolves towards equilibrium. The steady state is a constant column vector, $\chi_{0;i}= 1$ for all $i$. Hence, it is clear from \Eqs{lam+} and \eqref{lam+P01} that $\lam_0=0 $ for the steady mode.

\section{Solution selected by continued fraction}

In the eigenvalue equation \eqref{eprob0}, when any two of $\phi_0$, $\phi^+_1$ and $\phi^-_1$ are chosen independently, the rest of the $\phi^\pm_i$ can be determined from the subsequent equations in \eqref{eprobn}. Therefore, this set of equations should have two independent solutions. However, when we write \Eq{eprob0} in terms of the ratio $F^\pm_1=\phi_0/\phi^\pm_1$, the continued fraction method gives unique $F^\pm_1$, and therefore $\phi^+_1 $ and $\phi^-_1$ are no longer independent once $\phi_0$ is given. This shows that the continued fraction method yields only one of the two possible solutions. We will now show that the solution that is not produced by the continued fraction method can be divergent at large $\pb$. Therefore, the method automatically discards the unphysical solution.

For large $i= N\gg 1$, we can use \Eq{Fteq2} to solve for $F^\pm_N$ by approximating $F^\pm_{N+1}\approx F^\pm_N$. We then use \Eqs{xy} and \eqref{r} to obtain (omitting the $\pm$ superscript and $\p0$ for simplicity)
\begin{align}   \label{FN}
    F_N &\approx \frac{1}{2} ( {x_N}-\lb r_N) \pm \frac{1}{2} \sqrt{(x_N-\lb r_N)^2+4 y_N}\no
            &\approx \frac{1}{2} (1-\lb) e^{2\btb|{p}_{N}| }
            \left[1\pm \sqrt{1-\frac{4 e^{-2\btb|{p}_{N}| }}{(1-\lb)^2}}\right]\,.
\end{align}
The vanishing exponential in the square root allows us to express the square root as a series of polynomials. As a result, the two solutions are
\begin{align}   \label{F12}
    F^{(1)}_N &= (1-\lb) e^{2\btb|\pb_N| }\,, \\
     F^{(2)}_N &=\frac{1}{1-\lb} \,.
\end{align}
$F^{(1)}_N$ is just the boundary condition for $F^\pm_N$ we have made use of, see \Eq{Ftinfty}.

Using \Eq{phiN}, the first solution is
\begin{align}   \label{phi12}
    \phi^{(1)}_N&= \frac{e^{-2\btb|\pb_N|} }{1-\lb} \phi_{N-1}=\cdots = \left( \prod_{i=1}^N \frac{e^{-2\btb|\pb_i|} }{1-\lb}\right) \phi_0  \,,
\end{align}
which vanishes rapidly for large $N$. It is known that continued fraction method selects the solution with the behavior that as $N$ increases, $\phi_N$ either decreases in the fastest way or increases in the slowest way \cite{Risken}. $\phi^{(1)}_N$ is in fact the solution selected by the continued fraction method, and is consistent with the vanishing requirement of the eigenfunction at infinity. On the other hand, the second solution is
\begin{align}   \label{phi12b}
    \phi^{(2)}_N&= (1-\lb) \phi_{N-1}=\cdots=(1-\lb)^N  \phi_0 \,.
\end{align}
For $\lb>1$, $\phi^{(2)}_N$ can be a large quantity and is not consistent with the requirement that the eigenfunction should vanish at infinity. This solution is not produced by the continued fraction method.

From \Eq{phi12}, we also learn that if we are looking for eigenvalue that is close to $\lb = 1$, we need to iterate the continued fraction even more times (larger $N$), so that the approximation we used to expand the square root in \Eq{FN} remains valid. On the other hand, one may suggest using the ratio $\bar{F}^\pm_i =1/F^\pm_i$ for iteration in \Eqs{eprob0} or \eqref{ep10}, and \eqref{eprobn}. However, it is found that solution in terms of $\bar{F}^\pm_i$ produces only the eigenvalues with even indices. This is because for the solutions that correspond to the eigenvalues with odd indices, $\phi_0$ is either exactly zero or close to zero. The ratio $\bar{F}^\pm_1=\phi^\pm_1/\phi_0$ is therefore a divergent or very large quantity. The continued fraction method does not generate solutions for this situation.

\section{Reflection Symmetry of $\K$}
\label{sec-sym}

In this appendix we establish a reflection symmetry between the matrices $\K(\p0)$ and $\K(-\p0)$, which leads to a relationship between $\phi_\mu(\pb)$ and its mirror reflection $\phi_\mu(-\pb)$.

We start by defining a $\times$-operation that acts on a square matrix $\textbf{A}$, and a column vector $\psiv$ in the following ways,
\begin{align}   \label{xop}
    A^\times_{i,j}&\equiv A_{-i,-j} \,, & \psi^\times_i &\equiv \psi_{-i}\,.
\end{align}
This operation shuffles the components of the matrices, but leaves the values of the components intact.

A closer look at \Eq{eprobn} shows that it depends only on the magnitude of $\pb_{0;i}$. Furthermore, from \Eq{p0pm}, we have the relation
\begin{align}   \label{p0-}
    (-\p0)_{;i} = (-1)^i(-\p0-2i)&=-(-1)^i(\p0+2i)\no
    &=-(\pb_{0;-i})\,,
\end{align}
which implies
\begin{align}   \label{po-}
    |(-\p0)_{;i}| &=|\pb_{0;-i}|\,.
\end{align}
Based on these facts, we find that
\begin{align}   \label{K0xcomp}
    ({K^\times})_{i,j}(-\p0) &= {K}_{-i,-j}(-\p0)={K}_{i,j}(\p0)\,,
\end{align}
which is equivalent to
\begin{align}   \label{K0xK0}
   \K^\times(-\p0)=\K(\p0)\,.
\end{align}

Now we carry out the $\times$-operation on the eigenvalue equation \eqref{eprob} for $-\p0$. Using \Eq{K0xK0}, we obtain
\begin{align}   \label{eprobx}
    \K(\p0)\cdot \phiv^\times_\mu(-\p0)&=- \lb_\mu(-\p0) \phiv^\times_\mu(-\p0)\,.
\end{align}
On the other hand, the components of $ \phiv^\times_\mu(-\p0)$ is
\begin{align}   \label{phivxcomp}
    \phi^\times_{\mu;i}(-\p0)= \phi_{\mu;-i}(-\p0)=\phi_\mu[(-\p0)_{;-i}]=\phi_\mu(-\pb_{0;i})\,,
\end{align}
where we have used \Eq{p0-} in the last equality. In component's form, \Eq{eprobx} becomes
\begin{align}   \label{eprobxcomp}
    K_{ij}(\p0)\phi_\mu(-\pb_{0;j})&=-\lb_\mu(-\p0) \phi_\mu(-\pb_{0;i})\,.
\end{align}
Therefore, $\phi_\mu(-\pb_{0;i})$ is the eigenvector of $\K$ with eigenvalue $\lam_\mu(-\p0)$.
Comparing this with the component's form of the right eigenvalue problem \eqref{eprob},
\begin{align}   \label{eprobxR}
    K_{ij}(\p0)\phi_\mu(\pb_{0;j})&=-\lb_\mu(\p0) \phi_\mu(\pb_{0;i})\,,
\end{align}
and using the fact that the eigenvalues are real, non-negative and non-degenerate, we can match the eigenvalues according to
\begin{align}   \label{lb-p0}
    \lb_\mu(-\p0)=\lb_\mu(\p0)\,.
\end{align}
Furthermore, $\phi_\mu(-\pb_{0;i})$ is related to $\phi_\mu(\pb_{0;i})$ up to a phase. Since $\K$ and the eigenvectors are real, the phase can only take the real value $\pm 1$. We then conclude that
\begin{align}   \label{phixphivcomp}
    \phi_\mu(-\pb_{0;i})=\pm \phi_\mu(\pb_{0;i})\,.
\end{align}

This result enables us to construct a function $\tilde{\phi}_\mu(\pb)$ in Section \ref{sect-eigev} that is piecewise continuous in $\pb$, and is either symmetric or anti-symmetric with respect to mirror reflection in $\pb$.

\end{document}